# Theory MOND in a Friedmann-Robertson-Walker Cosmology as alternative to the Nonbaryonic Dark Matter paradigm

Nelson Falcón Universidad de Carabobo Dpto. de Física FACYT. Venezuela
nelsonfalconv@gmail.com

Modified Newtonian Dynamics (MoND) is an empirically modification of Newtonian gravity at largest scales in order to explain rotation curves of galaxies, as an alternative to nonbaryonic dark matter. But MoND theories can hardly connect themselves to the formalism of relativistic cosmology type Friedmann-Robertson-Walker. The present work posits the possibility of building this connection by postulating a Yukawa-like scalar potential, with non gravitational origin. This potential comes from a simple reflection speculate of the well–know potential of Yukawa and it is intended to describe the following physics scenarios: null in very near solar system, slightly attractive in ranges of interstellar distances, very attractive in distance ranges comparable to galaxies cluster, and repulsive to cosmic scales. As a result of introducing this potential into the typical Friedman equations we found that the critical density of matter is consistent with the observed density (without a dark matter assumption), besides this, MoND theory is obtained for interstellar scales and consequently would explain rotation curves. Also it is shown that Yukawa type inverse does not alter the predictions of the Cosmic Microwave Background neither the primordial nucleosinthesys in early universe; and can be useful to explain the large-scale structure formation.

**1. Introduction**.

The scientific cosmology is based on the description of gravitation by the General Theory of Relativity, more specifically in the solutions of the Friedmann equations for a model of isotropic and homogeneous universe on a large scale (metric Friedmann-Robertson-Walker FRW) and continued expansion according to Hubble's Law. According to this description the dynamics of the universe would be determined by the amount of matter existing in it, which in turn determines the large-scale geometry of space-time.

Recent observations of relic radiation in the Cosmic Microwave Background (CMB) have confirmed in essence the predictions of the Big Bang model and seem to corroborate the predictions of the FRW universe models with zero curvature (k = 0). More refined measures of inhomogeneity in the CMB (Hinshaw et al 2009, Komatsu et al 2010) with the measurements of the supernovae (SNe Ia) at high redshift (Riess et al, 1998, Pemlutter et al 1999) suggest the existence of a cosmic acceleration in accordance with the predictions of the model universe with constant cosmological ($\Lambda \neq 0$). This large-scale cosmic acceleration, also called dark energy, is today one of the most important enigmas of modern cosmology (Peebles Rastra 2003; Peebles 2007, Carroll Press 1992, and references therein) . Even more disturbing is the apparent contradiction between the models of the universe with zero curvature (k = 0) in the FRW formalism and the total density parameter ($\Omega$), according to which should be exactly equal to unity, but the observed density of matter is an order of magnitude lower than expected to play the null curvature (see Overduin and Wesson, 2008 and references therein).

Assuming that the dynamics of the universe is prescribed only by the force of gravity (as the only fundamental force astronomical scale) we encounter serious difficulties in describing the behavior of the Universe:

-i-Can not explain the rotation curves of galaxies, which show its incompatibility with the virialized mass of galaxies. (i.e. Sofue et al 1999, Sanders and McGaugh 2002; and references therein)

-ii-In rich clusters of galaxies, the mass observed in the form of stars and the gas mass inferred from the X-ray diffuse emission is significantly less than that required to maintain these systems gravitationally stable. ( Shirata et al 2005 and references therein)

-iii-A cosmological scales the observed baryonic matter density is much lower than predicted by the FRW models with cosmological constant and zero curvature. ( i.e. Overduin and Wesson 2008, and references therein)

The problem of missing mass and appears to affect the dynamics at all length scale beyond the Solar System (Freese 2000 and references therein). One solution has been to propose the requirement of missing material of unknown origin (non baryonic Dark Matter) with equally unknown properties and only interacts gravitationally with ordinary matter. However, after more than a decade of strenuous efforts: theoreticals, astronomical observations and laboratory experiments, only their existence has been suggested conjectural or paradigmatically.

In recent years there have been several alternatives to dark matter paradigm to explain the rotation curves of galaxies or as alternatives to the TGR and the Big Bang cosmology. Among the first calls include MOND theories, which reproduce successfully despite rotation curves of galaxies (i) can not resolve in his formalism the lack of dark matter at scales of galaxy clusters (ii) cosmological scale and (iii) It is based on the modification of the Gravitation Universal Law to scales larger than the solar system, where the dependence of the force of gravity does not necessarily verify the law of the inverse of the square. In the same vein are the Theories of Moffat (2005), who postulates a modification of the Universal Gravitation constant, although it is a very promising alternative to the paradigm of Dark Matter, faces problems that a variation of 25% or more in the constant acceleration of gravity would imply an abundance of helium incompatible with the observations (Reeves 1994 and references therein , also see Melnikov 2009 to constrain the variation of fundamental constants). Among the proposed changes to the TRG the most promising are those that postulate the existence of an additional scalar field metric tensor (Branks-Dicke theory) (Fujii and Maeda 2004) and within this so-called quintessence scenarios (Martin, J. 2008) to postulate a new fundamental interaction, additional gravity and electroweak interactions (electromagnetic and weak nuclear force) and strong nuclear force.

While it is true that Newton's law of gravitation, the inverse square law, has been highly supported in the laboratory to precisions greater than $10^{-8}$ for Eötvös-type experiments and spacecraft satellites (for comprenssible review see Gundlach 2005; Silverman 1987) there is no experimental evidence to confirm the validity of Newtonian dynamics beyond the Solar System. For a review of the many theoretical speculations about deviations from the $r^{-2}$ law see (Adelberger et al., 2003). Also Terrestrial and solar system experiments are shown to severely constrain the strength of an antigravity field with range much greater than 1 AU (Goldman 1987). From the very beginning of the pre-relativistic cosmology and ideas were raised on the modification of gravity at beyond the Solar System (i.e. Seeliger 1895; Bondi 1970)..

In relation to the hypothesis of non baryonic Dark Matter the history of science has shown many examples of local paradigmatically made to explain the behavior of nature that were then non-existent and replaced by alternatives measurable. Such as the cycles and epicycles of Ptolemy, the ether before the advent of the Special Theory of Relativity, the "caloric" (as elementary substance) before the work of Joule and Carnot. In all cases a review of the assumptions made in the phenomenological description of the processes led to a breakthrough in our understanding of natural reality.

You can apply for a modification to the theory of gravitation Newton (MOND) or the General Relativity, as Brank-Dicke (quintaessence scenarious), but would remain some formalism to connect these ideas (Milgrom, 2001; Sanders and McGaugh 2002) with the usual formalism and FRW models and the observables of the Big Bang model, such as acoustic peaks in the CMB fluctuations, the density of matter, the age of the universe in terms of the Hubble constant, primordial nucleosynthesis (baryogenesis) and the formation of structures from the primordial fluctuations.

In that vein we propose as an alternative to release the assumption that modern cosmology is based, namely that the dynamics of stars, galaxies and clusters of galaxies is determined solely by the force of gravity.

To this end we postulate the existence of a new fundamental interactions, whose origin is baryonic matter, similar to gravity and which acts differently at different length scales, as did the approach of Yukawa for the strong interaction (Yukawa, 1935); according to which nuclear power would be void, attractive or repulsive at different length scales. This new interaction, which we call here Inversed Yukawa Field (IYF) is built by the specular reflection of the Yukawa potential (Section 2), resulting in a cosmological constant depending on the comoving distance, ie a potential null near the Solar System, in agreement with the terrestrial experiments, weakly attractive to tens of kiloparsec scales consistent with MOND, strongly attractive at scales of tens of Megaparsec and repulsive cosmological scales of the order or greater than 50 Mpc in agreement with the cosmological constant.

We propose to show that this potential (IYF), built heuristically leads to a standard FRW model with nonzero cosmological constant, in which the density of matter observed is sufficient to verify the full (k = 0) without non baryonic dark matter hypothesis. In section 3 we will show that a quintessence of the type proposed (IYF) would be concomitant with the observations of CMB primordial baryogenesis and solve the horizon problem without the inflationary scenario, and solve the problem of exponential growth of large-scale structures the Universe. Finally the short dicussion and conclusions are shown in the lasts sections, 4 and 5 respectively.

## 2. FRW model with Inversed Yukawa field (IYF)

We assume that any particle with nonzero rest mass is subject to the Newtonian gravitational force by the law of Universal Gravitation, and an additional force that varies with distance, we call the Inverse Yukawa field. Also, without changing the argument, could be thought that the force of gravitation is bimodal (bigravity): varies as the inverse square of distance scales negligibly small compared to $r_0$, but it varies from very different when the comoving distance is about the kiloparsec or more. In this sense, our story line is a MOND Theory. It is also clear that the origin of this IYF is the baryonic mass like the Newtonian gravitational force.

This potential type Yukawa inverse per unit mass, is build starting from a reflection to speculate of the potential of Yukawa: null in very near solar system, slightly attractive in ranges of interstellar distances, very attractive in distance ranges comparable to galaxies cluster and repulsive to cosmic scales:

$$U(r) \equiv U_0(M)(r - r_0) e^{-\alpha/r} \qquad (1)$$

Where $U_0(M)$ is the magnitude of this potencial per unit mass (in units of N/kg) as a function of baryonic mass that causes the field, and $r_0$ is the orden of $50h^{-1}$ Mpc. Where h is the parameter Hubble, defined as $H_0 = 100 h\, km/s\, Mpc$ for the Hubbble constant at the present epoch.
.
Also, α is an coupling constant in order of $2.5h^{-1}$ Mpc. The figure 1 show the variation of $U/U_0$ respect to adimensional variable $x \equiv r/r_0$.

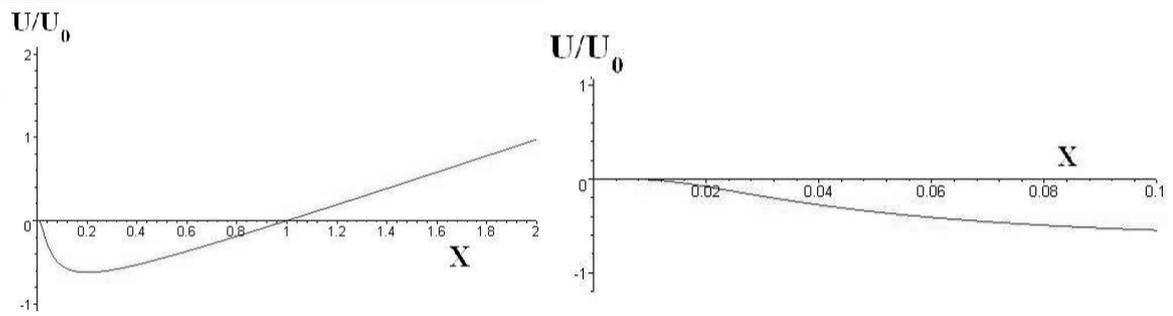

**Figure 1**: Potential Yukawa Inverse per unit mass as fuction of adimenssional comovil scale $x = (r/50h^{-1})Mpc$ ( Large-scale variation left, right: the function near the origin)

We can see that in scales of distance the order of Solar System this contribution is null, this is mildly attractive potecial distances of the order of the kiloparsec, strongly attractive at distances comparable to megaparsec and repulsive to cosmological scales.
Thus de Yukawa inversa force per unit mass, namely bimodal complement large-scale Newtonian gravitation would be:

$$F_{YI}(r) \equiv -\frac{U_0(M)}{r^2} e^{-\alpha/r}\left(r^2 + \alpha(r - r_0)\right) \qquad (2)$$

Also in the weak field approximation (x≪1) the Yukawa inverse Force per unit mass is given by:

$$F_{YI}(r \ll r_0) \approx -\frac{U_0(M)\,\alpha\, r_0}{r^2} \qquad (3)$$

But if x→0 this Force per unit mass is null, in accordance to the measures in experiments on Earth.
Now, we prove that (2) recovers the MOND Milgrom assumptions, is this for a given mass M, the asymptotic acceleration at r (in order to kiloparsec) goes as $r^{-1}$ (Milgrom, 2009), Thus r ≈ kpc then

$$\left|F_{YI}(r \ll r_0)\right| \approx \frac{U_0(M)r_0}{2r + \alpha} \approx \left(\frac{U_0(M)r_0}{2}\right) r^{-1} \qquad (4)$$

This expression represents the right side of the decrease of IYF, see figure 2; Note that the maximun occurs at a distance of the order to $x=10^{-2}$. Futhermore the Milgrom Theory can be to explain completely the rotation curves of galaxies, then IYF can be equaly can do. Remember that the usual Newton law for the gravitation adds to this force per unit mass (4).

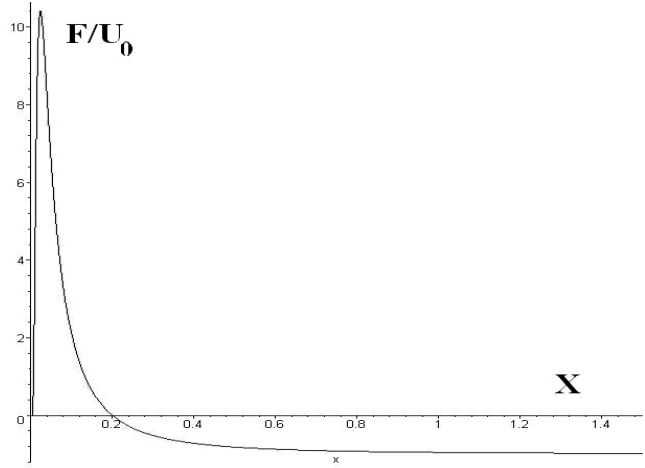

**Figure 2**: Force Yukawa Inverse per unit mass, as fuction of adimenssional scale
$$x = (r/50h^{-1})Mpc$$

Also in at cosmological range of comovil distance, the force Yukawa inverse is constant and provides the asymptotic cosmic acceleration. The minimun value of the potential occurs for

$$r_c = \frac{\alpha}{2}\left(\sqrt{1+\left(\frac{4r_0}{\alpha}\right)}-1\right) \quad (5)$$

If we assume, as before, that $\alpha \approx 2.5h^{-1}$ Mpc and $r_0 \approx 50h^{-1}$ Mpc, substituting in (2) then the maximun value of the force occurr in $r \approx 1.2\ h^{-1}$ Mpc, it`s the order to tipycal Abell radius of the clusters of galaxies. The assumption is justified because $r_0$ because it is the average distance between clusters of galaxies (i.e. see Guzzo 2002, and references therein). And alpha is calculated by (5) for the average value of almost smooth transcision distribution of galaxies to strong agglutination, in the order to $r_c = 10h^{-1}$ Mpc (i.e see Peebles and Ratra 2003, and references therein)

Let us now cosider a usual homogeneous and isotropic FRW metric with the line element:

$$ds^2 = g^{\mu\nu}g_{\mu\nu} = c^2dt^2 - R^2(t)\left(\frac{dr^2}{1-kr^2} + r^2d\theta^2 + r^2\sin^2\theta\,d\phi^2\right) \quad (6)$$

Where R(t) is the rate of the expansion, $g^{\mu\nu}$ is the metric tensor and k=0,-1,+1 is the scalar curvature for flat,open and closed universe, and c is the speed of light. Also consider a usual energy-momentum tensor for a perfect fluid as:

$$T^{\mu\nu} = (\rho + P)u^\mu u^\nu - p\,g^{\mu\nu} \quad (7)$$

Now, we assumet that $\Lambda \equiv \Lambda(r)$, it is the "force" cosmological, as dynamic variable, respect to the comovil distance. Without loss of generality, we can write: $\Lambda \equiv \Lambda(r) \propto U(r)$, then

$$\Lambda(r) \equiv \Lambda_0 \ (x-1)\ e^{-\alpha_0/x} \tag{8}$$

Where $\Lambda_0$ is a coupling constant of with dimensions of inverse square meters, and $\alpha_0 = 1/20$ is a dimensionless constant or $\alpha_0 = \alpha/r_0$. As before $x \equiv r/r_0$. Then $\Lambda_0 \approx 39\ H_0^2/c^2$ or $\Lambda_0 \approx 0.45\ h^2\ 10^{-50} m^{-2}$.

Thus the Einstein equation with cosmological term is given by:

$$\Re^{\mu\nu} - \frac{g^{\mu\nu}}{2}\Re + \Lambda(r) g^{\mu\nu} = \frac{-8\pi G}{c^2} T^{\mu\nu} \tag{9}$$

Where $\Re^{\mu\nu}$ is the Riemann tensor and $\Re$ is the Riemann scalar. It`s easy see that the cosmological term leads to usual Friedmann equations (repeating the usual derivation of the equations of Friedmann, as in Weimberg (1972) or Adler, Bazin and Schiffer (1965), but in the assumption that $\Lambda \equiv \Lambda(r)$):

$$\left(\frac{\dot{R}(t)}{R(t)}\right)^2 + \frac{kc^2}{R^2(t)} = \frac{8\pi G}{3}\rho + \frac{\Lambda(r)c^2}{3} \tag{10}$$

$$\frac{2\ddot{R}(t)}{R(t)} + \left(\frac{\dot{R}(t)}{R(t)}\right)^2 + \frac{kc^2}{R^2(t)} = -\frac{8\pi G}{c^2}P + \Lambda(r)c^2 \tag{11}$$

### 3. Cosmological Consequences

But the definition of the critical density change because the potential Yukawa in now nonzero when k=0. The critical density ($\rho_c$), using (10) is now:

$$H_0^2 \equiv \left(\frac{\dot{R}(t)}{R(t)}\right)^2 = \frac{8\pi G}{3}\rho_c + \frac{\Lambda(r=r_c)c^2}{3} \tag{12}$$

Thus the Friedmann equations (10)(11) are:

$$\frac{kc^2}{R^2(t)} = H_0^2\left[\Omega_m\left(1 - \frac{c^2\Lambda(r=r_c)}{3H_0^2}\right) + \Omega_\Lambda(r) - 1\right] \tag{13}$$

$$q_0 = \frac{\Omega_m}{2}\left(1 + \frac{c^2\Lambda(r=r_c)}{3H_0^2}\right)^{-1}(1 + \frac{3P}{c^2\rho}) - \Omega_\Lambda(r) \tag{14}$$

Where we used the standars notation (Peacock 1999): $\Omega_m \equiv \rho/\rho_c$, $\Omega_\Lambda \equiv \Lambda c^2/3H_0^2$, $q_0 \equiv -\frac{\ddot{R}}{R}H_0^{-2}$ for the dimensionless density parameter of matter, dimensioless cosmological parameter and deceleration parameter respectively. Notice however that $\rho_c$ is given now by (12).

It is easy to see that (13) and (14) are the usual Friedmann equations when the Yukawa inverse potential is null. Using, as before, $\Lambda_0 \approx 39\, H_0^2/c^2$ alongside (5) and (8) we obtain

$$\Lambda(r = r_c) \cong -24.3 \frac{H_0^2}{c^2} \qquad (15)$$

The (12) and (15) follows

$$\rho_c = \frac{3H_0^2}{8\pi G}\left(1 - \frac{\Lambda(r = r_c)c^2}{3H_0^2}\right) \cong 9.1 \frac{3H_0^2}{8\pi G} \approx 2.53\ 10^{12} h^2\, M_{sun}/Mpc^3 \qquad (16)$$

It is clear that the critical density value increases, because the critical mass has been underestimated by the usual definition, the Yukawa field, must join the mass equivalent to the energy of the field. have to take into account. The mean value of central density value in core of the clusters of galaxies is $3\ 10^{15}\, M_{sun}/Mpc^3$ ( Jones and Forman 1984).
Notice that, if we define:

$$\Omega_{YIF} \equiv -\frac{\Lambda(r = r_c)c^2}{3H_0^2} \qquad (17)$$

And with (15), we obtain $\Omega_{YIF} \cong 8.1$
The Freedman equation (13) and (14) are as:

$$\frac{kc^2}{R^2(t)} = H_0^2\left[\Omega_m(1+\Omega_{YIF}) + \Omega_\Lambda - 1\right] \qquad (18)$$

$$q_0 = \frac{\Omega_m}{2}(1 - \Omega_{YIF})^{-1}(1 + \frac{3P}{c^2\rho}) - \Omega_\Lambda \qquad (19)$$

But now $\Omega_\Lambda$ is a dynamic parameter, using (8) should be evaluated at cosmological scales:

$$\Omega_\Lambda \equiv \Omega_\Lambda(r \gg r_c) \equiv \Lambda_0 \frac{c^2}{3H_0^2}(x-1)\, e^{-\alpha_0/x} \qquad (20)$$

Thus for scales larger distances such as to ensure homogeneity and isotropy, namely at scales larger than 50h$^{-1}$ Mpc or $x \equiv \frac{r}{r_c} \gg 1$. Then the behavior of the $\Lambda$ function is asymptotic (see figure 1 by x>>1) and can be estimated as:

$$\Omega_\Lambda \cong \Lambda_0 \frac{c^2}{3H_0^2} \alpha_0\, (x-1) \qquad (21)$$

Reeplacing (15) into (21) with x≈2, equivalent to a comoving distance of the order of 100h$^{-1}$Mpc, range for which the galaxies behave as particles in accordance with a FRW model and the assumptions of homogeneity and isotropy. we obtain:

$$\Omega_\Lambda \cong \frac{39}{3}\frac{1}{20}(2-1) \approx 0.65 \qquad (22)$$

very close to the usual value 0.7 (i.e. Freedman 2000, and references therein)
Now, the import result is that k=0 and $\Omega_{YIF} \neq 0$ does not required the nonbaryonic dark matter assumption. I.e using (15) and $\Omega_\Lambda \approx 0.7$ we obtain $\Omega_m \approx \Omega_b = 0.03$ as the typical value for a flat universe model without nonbaryonic dark matter.
    For the early stages of the evolution of the Universe, we can find from equations (13) and (14) the relationship between the scale factor R (t) and the state variables ρ and P; derivative (13) to replace $d^2R/dt^2$ in (14) and using again (13) we obtain, as usual

$$-\frac{3dR(t)}{R(t)} = \frac{d\rho}{\rho + P/c^2} \quad (23)$$

Therefore, the dependence of thermodynamic variables density and pressure on the scale factor R (t) remain unchanged and may, as usual, used the state equation $p = \omega\rho c^2$ for radiation (ω=1/3), dust (ω=0) and vacuum (ω= -1)(Peacock 1999, Peebles 1993). Similarly neither affects the calculation of time decoupling between matter and radiation.

The Mattig formula (Mattig 1959, Dabrowski and Stelmach 1986) only modified by introducing the term critical density. Using (18) and (21), The Freedmann- Lemaitre equation (by flat universe) is now:

$$H^2 = H_0^2 \left[ \Omega_m (1+z)^3 + \Omega_\Lambda \right] \quad (24)$$

has been omitted for simplicity the contribution of the radiation density, but can be incorporated as a sum ,multiplied by the factor $(1 + z)^4$ , without loss of generality in the discussion. By flat model (k=0) we can write the limit the age of the universe at redshift z as:

$$\tau = H_0^{-1} \int_0^\infty \left[ (1+z)^3 \Omega_m (1+\Omega_{YIF}) + \Omega_\Lambda \right]^{-\frac{1}{2}} \frac{dz}{z+1} \quad (25)$$

Note that if the Yukawa field is zero we obtain the standars equation for the age of the universe. in ΛFRW flat model. Also using (15) here, the limit of age of the universe increcrease in about 30%, until $17h^{-1}$ Gy, because Ωm increase a factor ten.

Remember the cosmic age problem associated with oldest globular clusters in galactic halo and the quasar APM 08279 + 5255 at z = 3.91. If the age estimates of these objects are correct, the cosmic age puzzle still remains in the standard cosmology (Pont et al. 1998, Ma, et al 2009, Wang et al 2010, Yang and Zhang 2010), but not in the present model with Yukawa inverse Potential. Also Wang and Zhang (2008) suggested that the introduction of new interaction may be helpful to remove the cosmic age problem and demonstrated that the dark energy paradigme alone cannot remove the high-z age problem.

For the other hand the formation of large-scale structures should be reviewed in the context of a theory of quintessence, as in the case of the inclusion of the inverse Yukawa potential. Since the length of Jeams measuring the dimensions of stability of protogalactic clouds in the form (Jeans, 1928)

$$\lambda_J \approx \tau_g c_s = \tau_g \sqrt{\frac{dP}{d\rho}} \quad (26)$$

Where $c_s$ is velocity of sound and $\tau_g$ is the free ball time.

To proto stellar clouds, the time of free fall is only the Newtonian gravity, but may not be equal to dimensions higher during times of the early universo, where the dimensions of the clouds would have colossal (proto galaxies) to megaparsec scales. At such scales would have bigravity (Rossi 2009, Blas 2006), and the inverse Yukawa potential per unit mass would have to join the force of gravitation. So

$$\lambda_J = \tau_g c_s \propto \frac{c_s}{\sqrt{G\rho + F_{YI}(r)}} = \frac{c_s}{\sqrt{G\rho + m\frac{U_0(M)}{r^3} e^{-\alpha/r}\left(r^2 + \alpha(r-r_0)\right)}} \quad (27)$$

Where we used (2) as expresion of the Yukawa inverse force per unit mass. Notice that the additional term is exponential , and could be an interesting approach in the hierarchical fragmentation, also "there exists no convicing theory of galaxy formation". ( i.e. Ross 1997 and references therein).

The decrease in the Jeans length scale implies that fragmentation could begin in earlier times and consequently favors the formation of proto-galaxies from primordial clouds, whose free fall time decreases. Also the growth of structure depends linear matter density constrast and their description is different a diferent scales (i.e. interesting paper Bueno-Sanchez et al 2010) as would be expected for a dynamic cosmological term $\Lambda(r)$. We wish to emphasize the function of linear growth of density perturbations in a flat universe model with cosmological constant has been reported by Eisenstein (1997) . If instead of using a constant expression as in (23), is used as a dynamic term (8) would be obtained different functions of growth of disturbances at different length scales.

It should be noted that, under or intergalactic scales, the dimensions of comoving distances are such that $r \ll r_c$ and the term of the inverse Yukawa force is negligible and thus not expected in this case variations in the fragmentation of clouds by the active regions of star formation.

## 4. Discussion

The proposed IYF is proportional to the baryonic mass, through the coupling constant $U_0(M)$. The particles with zero rest mass as photons would not be affected and therefore not expected variations in the CMB. Because of this IYF potential, unlike the force of gravity on the context of general relativity, not cause a curvature in space-time, in full accord with the hypothesis of Einstein (1916) to the propose the cosmological constant, independent of Riemann tensor as in equation (9). The discussion of the results of the measurements of the CMB anisotropies should be reviewed if the Friedmnn equations include a term that represents the bigravity as $\Omega_{YIF}$ as in the expressions (18) and (19).

We can look that the Big Bang nucleosythesis takes place in the early universe, and the baryogenesis calculation (see Reeves 1994 for comprenssible review, Burles et al 2001, Steigman 2010 and references therein) explicitly used Eq. (23) which is identical with or without YIF. Other models (Moffat 2006) include scalar fields modified in Newton's constant of gravitation (G) could be in conflict with primordial nucleosynthesis, which explicitly uses this value compared to the Fermi constant to account for the abundances of light elements observed in the universe. Remark that not the case in bigravity type models, MoND theory and/or YIF as presented here, for amending the law of gravitation only comoving distance scales forty orders of magnitude higher than the average distance per nucleon in the primordial plasma.

Another interesting controversy in recent years is that concerning to anomalous acceleration from de Pioneer 10/11 spacecraft when traveling through the outer reaches of the solar system. Indicated the presence of a small, anomalous, blue-shifted Doppler frequency drift, interpreted as a sunward acceleration of $a_p = (8.74 \pm 1.33) \times 10^{-10}$ m/s$^2$ (Anderson et al 1998, 2002). This signal has become known as the Pioneer anomaly; the nature of this anomaly is still being investigated (Toth 2009, Olsen 2007). Another possible interpretation of the Pioneer anomaly is to consider the bigravity, For example, if in addition Newtonian gravity, there is a counterpart of the Inverse Yukawa force, as in Equation (3). In this case the coupling constant $U_0(M) \equiv \xi\, M_{sum,}$ lest the IYF is proportional to tha mass that caused the field. But the equation (3) varies with the inverse square of the distance sun-pioneer, then we assument that average distance between 20 to 70 UA, i.e. $r \approx 45\ UA$, thus

$$a_p \approx \frac{(\xi M_{sun}) \alpha r_0}{r^2} \quad (28)$$

Obviously the left side contains the effective force caused by anomalous acceleration (subtracted the Newtonian gravitation acceleration of the sun), the right side Inverse Yukawa acceleration for the average value of distance (45 UA). We obtain $\xi \approx 3,337 \ 10^{-41}$ N/kg$^2$ and then $U_0(M) = \xi \ M_{sum} \approx 6.64 \ 10^{-11}$ m/s$^2$, value very close to the Universal gravitational constant (G) and the other units. This is a curious result for which we have no explanation because gravity is not taken into account and α and $r_0$ reflect the behavior of the universe at cosmological scale. This is certainly a very crude approximation but it`s easy to see in figure 2 that in the range of 20 to 40 UA the IYF varies very slowly and therefore its contribution to the acceleration is almost constant in this range of distance (very small compared with $r_0$).

On the other hand, the missing mass in clusters of galaxies identified by Zwicky (1933) to calculate the mass excess using the Virial theorem could easily be resolved without invoking non-baryonic dark matter. Indeed, apart from the Clausius virial expression (Goldstein 2000):

$$2\left\langle \frac{m}{2} v^2 \right\rangle = -\left\langle \sum_i \vec{F}_i \cdot \vec{r}_i \right\rangle \quad (29)$$

But now, the large scale interaction are the Newtonian gravitational force and the Yukawa inverse field, then:

$$\left\langle v^2 \right\rangle = \left\langle \frac{GM}{r} + \frac{U_0(M)}{r} e^{-\alpha/r} \left[r^2 + \alpha(r - r_0)\right] \right\rangle \quad (30)$$

Use of the Virial theorem, as eq. (29), with only Newtonian gravity law implies that the average mass of a star is just the order of the solar mass and average mass of the proto-galaxies is just the mass of the the Milky Way (i.e. see Gonzalo et al 1995) in open contradiction with our knowledge of the universe and the Copernican Principle, according to which our location is not special in any way. When considering a long-range gravitational interaction, as Yukawa inverse potential or another MoND, through (30) we reverse this paradox.

Also, Mach Principle say that the inertia of an object as the gravitational interaction of its gravitational mass with the distant matter distribution in the universe (Mach, E. 1893, Singer 2005). For some, the Mach principle, which outlines a local connection between inertia and mass distribution on a large scale of the Universe, is more philosophical than scientific because until now all attempts have failed their precise mathematical formulation. The TRG to connect with inertia gravitational fields through the field equations, seem to enter the Mach principle, but not fully verified to be established as the boundary conditions of the Field Equations (Bondi 1951). The MoND can be interpretation as a new connection between the universe at large and local inertia (Darabi, 2010). Based on this interpretation, the Yukawa Inverse field can be fully comply the Mach Principle , through of the incorporation of the dynamic cosmological term Λ(r), which depends explicitly of cosmological quantities α, $r_o$ and $r_c$ in the equation (1), (2), (3) and (4). Certainly it is possible that this cosmological term would also depend on time, in this case the connection with the Higg's global field can be easy through the cosmological term Equations (21) and (22) as following Sivaram (2009) or Sivaran and Arun.(2009) .

Ishak et al (2010) have shown that the cosmological constant contributes a factor of second order in the angle of deflection from gravitational lenses. It is clear that including the effect of IYF through (21) and (22) also lead to a prediction in the variations of the estimated mass of gravitational lenses using the same estimates of Ishak (2008).

The origin of the scalar field proposed is beyond the scope of this work, we must bear in mind that there is not still a quantum theory of gravity, or have been directly detected gravitons (except perhaps PSR B1913+16 binary pulsar measurements that could be taken as indirect evidence of gravitational waves). There is controversial evidence about the origin of quantum and gravitational phenomena in the literature that are still open and that could justify the existence of a Yukawa type field as proposed (Raut and Sinha1981 Burgess and Cloutier 1988), also "The existence of an intermediate range coupling to the baryon number or hypercharge of the materials was confirmed." (Fischbach et al 1986) and Bezerra et al (2010) report "stronger constraints on the parameters of Yukawa-type corrections to Newtonian gravity from measurements of the lateral Casimir force" (see review by Decca et al 2009). The cosmological constant can be built through the extended action in the Palatini Lagrangian (Rosenthal 2009), so the variational formulation YIF field can be made by extension.

## 5. Conclusions

The traditional way to define the critical density deserves rethinking, under the action of a scalar field in addition to gravitation (in MoND type theories or IYF model discussed here) is unclear which corresponds to the density of matter in the Einstein-DeSitter Model. A change in the critical density, as in expression (12) opens up new theoretical posibiliades to avoid the paradigm of non-baryonic dark matter.

Have been proposed several modifications to Newtonian gravity, even operating on a large scale or type Yukawa potential (White and Kochanek, 2001; Amendola et al, 2004; Sealfon et al., 2005; Reynaud and Jaekel, 2005; Shirata et al., 2005; Sereno and Peacock, 2006, Moffat 2006, Berezhiani et al 2009). Clearly, the incompatibility between the flatness of the Universe ($k = 0$) and the density of matter comes from the Friedmann equation in its conventional form, which is removed if it is assumed that the Newtonian gravity act together to some scalar field (bigravity) or within a framework of Modified Newtonian Dynamics, maybe as like the Inverse Yukawa Potential proposed.

Regardless of whether the expression here proposed for the so-called Inverse Yukawa force per unit mass, is exactly the proposal here, we see the inclusion of a MOND theory expression through some form of dynamic cosmological term, ie a function of comoving distance as eq. (10) and (11), could be a viable alternative to the paradigm of non-baryonic dark matter and is concomitant with FRW cosmology. The cosmological constant then becomes a variable cosmological term . In such a scenario the low value of Λ merely reflects the fact the Universe is old . In general however this means modifying equations and/or introducing new forms of matter such as scalar fields .

Add a scalar field as the IYF proposed here, or MoND theory that corresponds to a kind of bigravity also imply that the masses of the nuclei of galaxies (Black Holes) have been overestimated, as well as masses as inferred by the gravitational lensing, since the scalar field as a summative term, contributes additively in the calculation of the gravitational potential. At large distances from the sources, the reduction in the Newtonian field with the inverse of the square would be offset by an interaction that is growing at much greater distances. These long-range interaction also could be caused by the baryonic mass and therefore would be calculable with physics usual.

**Acknowledgmeents**
The author thaks Dr. Williams Pitter (LUZ FEC Dpto Fisica, Maracaibo Venezuela) for useful discusions and sugerences.